\documentclass[aps,prl,twocolumn,superscriptaddress,showpacs,preprintnumbers,amsmath,amssymb]{revtex4-1}
\usepackage{graphicx} 
\usepackage{dcolumn}  

\graphicspath{{ps}}

\newcommand{\upsss}{\Upsilon(\text{4S})}
\newcommand{\upss}{\Upsilon(\text{2S})}
\newcommand{\ups}{\Upsilon(\text{1S})}

\newcommand{\A}{A^{0}}
\newcommand{\Amass}{M_{A^{0}}}
\newcommand{\DMmass}{M_{\chi}}
\newcommand{\Br}{{\cal B}}
\newcommand{\Gmass}{\text{GeV/}\textit{c}^{2}}

\begin{document}



\title{ \quad\\[1.0cm] Search for a light CP-odd Higgs boson and low-mass dark matter at the Belle experiment}

\noaffiliation
\affiliation{University of the Basque Country UPV/EHU, 48080 Bilbao}
\affiliation{Beihang University, Beijing 100191}
\affiliation{University of Bonn, 53115 Bonn}
\affiliation{Brookhaven National Laboratory, Upton, New York 11973}
\affiliation{Budker Institute of Nuclear Physics SB RAS, Novosibirsk 630090}
\affiliation{Faculty of Mathematics and Physics, Charles University, 121 16 Prague}
\affiliation{Chonnam National University, Kwangju 660-701}
\affiliation{University of Cincinnati, Cincinnati, Ohio 45221}
\affiliation{Deutsches Elektronen--Synchrotron, 22607 Hamburg}
\affiliation{Duke University, Durham, North Carolina 27708}
\affiliation{University of Florida, Gainesville, Florida 32611}
\affiliation{Key Laboratory of Nuclear Physics and Ion-beam Application (MOE) and Institute of Modern Physics, Fudan University, Shanghai 200443}
\affiliation{Justus-Liebig-Universit\"at Gie\ss{}en, 35392 Gie\ss{}en}
\affiliation{Gifu University, Gifu 501-1193}
\affiliation{II. Physikalisches Institut, Georg-August-Universit\"at G\"ottingen, 37073 G\"ottingen}
\affiliation{SOKENDAI (The Graduate University for Advanced Studies), Hayama 240-0193}
\affiliation{Hanyang University, Seoul 133-791}
\affiliation{University of Hawaii, Honolulu, Hawaii 96822}
\affiliation{High Energy Accelerator Research Organization (KEK), Tsukuba 305-0801}
\affiliation{J-PARC Branch, KEK Theory Center, High Energy Accelerator Research Organization (KEK), Tsukuba 305-0801}
\affiliation{Forschungszentrum J\"{u}lich, 52425 J\"{u}lich}
\affiliation{IKERBASQUE, Basque Foundation for Science, 48013 Bilbao}
\affiliation{Indian Institute of Science Education and Research Mohali, SAS Nagar, 140306}
\affiliation{Indian Institute of Technology Bhubaneswar, Satya Nagar 751007}
\affiliation{Indian Institute of Technology Guwahati, Assam 781039}
\affiliation{Indian Institute of Technology Hyderabad, Telangana 502285}
\affiliation{Indian Institute of Technology Madras, Chennai 600036}
\affiliation{Indiana University, Bloomington, Indiana 47408}
\affiliation{Institute of High Energy Physics, Chinese Academy of Sciences, Beijing 100049}
\affiliation{Institute of High Energy Physics, Vienna 1050}
\affiliation{INFN - Sezione di Napoli, 80126 Napoli}
\affiliation{INFN - Sezione di Torino, 10125 Torino}
\affiliation{Advanced Science Research Center, Japan Atomic Energy Agency, Naka 319-1195}
\affiliation{J. Stefan Institute, 1000 Ljubljana}
\affiliation{Institut f\"ur Experimentelle Teilchenphysik, Karlsruher Institut f\"ur Technologie, 76131 Karlsruhe}
\affiliation{Kavli Institute for the Physics and Mathematics of the Universe (WPI), University of Tokyo, Kashiwa 277-8583}
\affiliation{Kennesaw State University, Kennesaw, Georgia 30144}
\affiliation{Department of Physics, Faculty of Science, King Abdulaziz University, Jeddah 21589}
\affiliation{Korea Institute of Science and Technology Information, Daejeon 305-806}
\affiliation{Korea University, Seoul 136-713}
\affiliation{Kyoto University, Kyoto 606-8502}
\affiliation{Kyungpook National University, Daegu 702-701}
\affiliation{LAL, Univ. Paris-Sud, CNRS/IN2P3, Universit\'{e} Paris-Saclay, Orsay}
\affiliation{\'Ecole Polytechnique F\'ed\'erale de Lausanne (EPFL), Lausanne 1015}
\affiliation{P.N. Lebedev Physical Institute of the Russian Academy of Sciences, Moscow 119991}
\affiliation{Faculty of Mathematics and Physics, University of Ljubljana, 1000 Ljubljana}
\affiliation{Ludwig Maximilians University, 80539 Munich}
\affiliation{Luther College, Decorah, Iowa 52101}
\affiliation{University of Malaya, 50603 Kuala Lumpur}
\affiliation{University of Maribor, 2000 Maribor}
\affiliation{Max-Planck-Institut f\"ur Physik, 80805 M\"unchen}
\affiliation{School of Physics, University of Melbourne, Victoria 3010}
\affiliation{University of Mississippi, University, Mississippi 38677}
\affiliation{University of Miyazaki, Miyazaki 889-2192}
\affiliation{Moscow Physical Engineering Institute, Moscow 115409}
\affiliation{Moscow Institute of Physics and Technology, Moscow Region 141700}
\affiliation{Graduate School of Science, Nagoya University, Nagoya 464-8602}
\affiliation{Kobayashi-Maskawa Institute, Nagoya University, Nagoya 464-8602}
\affiliation{Universit\`{a} di Napoli Federico II, 80055 Napoli}
\affiliation{Nara Women's University, Nara 630-8506}
\affiliation{National Central University, Chung-li 32054}
\affiliation{National United University, Miao Li 36003}
\affiliation{Department of Physics, National Taiwan University, Taipei 10617}
\affiliation{H. Niewodniczanski Institute of Nuclear Physics, Krakow 31-342}
\affiliation{Nippon Dental University, Niigata 951-8580}
\affiliation{Niigata University, Niigata 950-2181}
\affiliation{University of Nova Gorica, 5000 Nova Gorica}
\affiliation{Novosibirsk State University, Novosibirsk 630090}
\affiliation{Osaka City University, Osaka 558-8585}
\affiliation{Pacific Northwest National Laboratory, Richland, Washington 99352}
\affiliation{Panjab University, Chandigarh 160014}
\affiliation{Peking University, Beijing 100871}
\affiliation{University of Pittsburgh, Pittsburgh, Pennsylvania 15260}
\affiliation{Theoretical Research Division, Nishina Center, RIKEN, Saitama 351-0198}
\affiliation{University of Science and Technology of China, Hefei 230026}
\affiliation{Showa Pharmaceutical University, Tokyo 194-8543}
\affiliation{Soongsil University, Seoul 156-743}
\affiliation{Stefan Meyer Institute for Subatomic Physics, Vienna 1090}
\affiliation{Sungkyunkwan University, Suwon 440-746}
\affiliation{School of Physics, University of Sydney, New South Wales 2006}
\affiliation{Department of Physics, Faculty of Science, University of Tabuk, Tabuk 71451}
\affiliation{Tata Institute of Fundamental Research, Mumbai 400005}
\affiliation{Excellence Cluster Universe, Technische Universit\"at M\"unchen, 85748 Garching}
\affiliation{Toho University, Funabashi 274-8510}
\affiliation{Department of Physics, Tohoku University, Sendai 980-8578}
\affiliation{Earthquake Research Institute, University of Tokyo, Tokyo 113-0032}
\affiliation{Department of Physics, University of Tokyo, Tokyo 113-0033}
\affiliation{Tokyo Institute of Technology, Tokyo 152-8550}
\affiliation{Tokyo Metropolitan University, Tokyo 192-0397}
\affiliation{Virginia Polytechnic Institute and State University, Blacksburg, Virginia 24061}
\affiliation{Wayne State University, Detroit, Michigan 48202}
\affiliation{Yamagata University, Yamagata 990-8560}
\affiliation{Yonsei University, Seoul 120-749}
\author{I.~S.~Seong}\affiliation{University of Hawaii, Honolulu, Hawaii 96822} 
\author{S.~E.~Vahsen}\affiliation{University of Hawaii, Honolulu, Hawaii 96822} 
  \author{I.~Adachi}\affiliation{High Energy Accelerator Research Organization (KEK), Tsukuba 305-0801}\affiliation{SOKENDAI (The Graduate University for Advanced Studies), Hayama 240-0193} 
  \author{H.~Aihara}\affiliation{Department of Physics, University of Tokyo, Tokyo 113-0033} 
  \author{S.~Al~Said}\affiliation{Department of Physics, Faculty of Science, University of Tabuk, Tabuk 71451}\affiliation{Department of Physics, Faculty of Science, King Abdulaziz University, Jeddah 21589} 
  \author{D.~M.~Asner}\affiliation{Brookhaven National Laboratory, Upton, New York 11973} 
  \author{V.~Aulchenko}\affiliation{Budker Institute of Nuclear Physics SB RAS, Novosibirsk 630090}\affiliation{Novosibirsk State University, Novosibirsk 630090} 
  \author{T.~Aushev}\affiliation{Moscow Institute of Physics and Technology, Moscow Region 141700} 
  \author{R.~Ayad}\affiliation{Department of Physics, Faculty of Science, University of Tabuk, Tabuk 71451} 
  \author{V.~Babu}\affiliation{Tata Institute of Fundamental Research, Mumbai 400005} 
% ...
\collaboration{The Belle Collaboration}


\begin{abstract}
We report on the first Belle search for a light CP-odd Higgs boson, $A^{0}$, that decays into low mass dark matter, $\chi$, in final states with a single photon and missing energy. We search for events produced via the dipion transition $\Upsilon(\textrm{2S})\rightarrow\Upsilon(\textrm{1S})\pi^{+}\pi^{-}$, followed by the on-shell process $\Upsilon(\textrm{1S})\rightarrow\gamma A^{0}$ with $A^{0} \rightarrow\chi\chi$, or by the off-shell process $\Upsilon(\textrm{1S})\rightarrow\gamma\chi\chi$. Utilizing a data sample of 157.3 $\times$ 10$^{6}$ $\Upsilon(\textrm{2S})$ decays, we find no evidence for a signal. We set limits on the branching fractions of such processes in the mass ranges $M_{A^{0}} <$ 8.97 $\textrm{GeV/}\textit{c}^{2}$ and $M_{\chi} <$ 4.44 $\textrm{GeV/}\textit{c}^{2}$. We then use the limits on the off-shell process to set competitive limits on WIMP-nucleon scattering in the WIMP mass range below 5~$\textrm{GeV/}\textit{c}^{2}$.
\end{abstract}


\maketitle

\tighten

{\renewcommand{\thefootnote}{\fnsymbol{footnote}}}
\setcounter{footnote}{0}

Identifying the nature of dark matter (DM) is a long-standing yet unsolved problem in astronomy and particle physics. DM may consist of weakly interacting massive particles (WIMPs), which are postulated in popular extensions of the standard model (SM)~\cite{Kane:2000ew}. Numerous experiments aim to directly detect WIMPs, but no clear evidence has emerged to date. WIMPs are generally expected to have masses in the 100 $\Gmass$ to 1 TeV/$c^{2}$ range, but there are also scenarios with DM particle masses below 100 $\Gmass$~\cite{Gunion:2005rw,Kaplan:1991ah,Kaplan:2009ag}. Such low mass DM particles, $\chi$, can be produced in interactions of SM particles through the exchange of a CP-odd Higgs boson $\A$~\cite{Gunion:2005rw, Petraki:2013wwa}, which is part of the Next-to-Minimal Supersymmetric Model (NMSSM)~\cite{Ellwanger:2009dp}. Searches for low mass DM particles from $\Upsilon$ decays at collider experiments have been discussed in Refs.~\cite{Fayet:2006sp,Yeghiyan:2009xc}. Hadron colliders would be insensitive to DM particles in the final state without additional constraints. However, at a B factory, such invisible decays can still be measured by utilizing the dipion transition $\upss\rightarrow\pi^{+}\pi^{-}\ups$, which enables us to tag $\ups$ mesons without reconstructing them. Since the mass of the $\A$ is unknown, we consider two processes: the on-shell process $\ups\rightarrow\gamma\A$ with $\A\rightarrow\chi\chi$; and the off-shell process
$\ups\rightarrow\gamma\chi\chi$. 
The SM process $\ups \rightarrow \gamma \nu\bar{\nu}$ has the same final state as the signal, but is predicted to have a branching fraction (BF) $\Br(\ups \rightarrow \gamma \nu\bar{\nu})$ of the order of $10^{-9}$~\cite{Yeghiyan:2009xc}, which is three orders of magnitude below our experimental sensitivity. The most stringent existing upper limits on the processes considered here were set by the BaBar experiment: $\Br(\ups \rightarrow \gamma \A)\times \Br(\A \rightarrow \chi\chi)$ $<$ ($1.9-37$)$\times10^{-6}$ for $\Amass <$ 9.0 $\Gmass$ and $\Br(\ups \rightarrow \gamma\chi\chi)$ $<$ ($0.5-24$)$\times10^{-5}$ for $\DMmass <$ 4.5 $\Gmass$~\cite{delAmoSanchez:2010ac}, both at 90$\%$ confidence level (C.L.).

This analysis uses a data sample with an integrated luminosity of 24.9~fb$^{-1}$, corresponding to (157.3~$\pm$~3.6)~$\times$~$10^{6}$ $\upss$ decays~\cite{Brodzicka:2012jm}, collected with the Belle detector~\cite{Belle} at the KEKB e$^{+}$e$^{-}$ asymmetric-energy collider~\cite{KEKB}. We generate one million Monte Carlo (MC) simulated events for each of the on-shell and the off-shell processes, the on-shell process with 20 different values of the $\A$ mass, $\Amass$, and the off-shell process with 10 different values of the $\chi$ mass, $\DMmass$. The $\A$ in the on-shell process is assumed to have zero spin while the off-shell process is modeled with a phase space distribution. In this paper, we do not assume a specific model for the $\A$ and $\chi$. The $\upss\rightarrow\pi^{+}\pi^{-}\ups$ transition is simulated using the EvtGen model to describe the decays of a vector particle to a vector particle and two pions~\cite{Lange:2001uf}.

Both signal processes produce only three detectable particles: two charged pions, which
have low transverse momentum; and a photon that deposits energy in the electromagnetic calorimeter (ECL). The small number of charged tracks and their low momenta are difficult to trigger on, thus the main level-1 (L1) triggers used are related to the ECL. Instead of using all possible L1 triggers in this analysis, we only use the two highest-efficiency L1 triggers, which require the total deposited energy in the ECL to be larger than 1.0 GeV with a cosmic ray veto applied and larger than 3.0 GeV without the veto. This choice reduces the systematic uncertainty on the L1 trigger efficiency. The signal trigger efficiency is a function of photon energy. In the energy region from 0.5 to 5 GeV, our L1 trigger requirements reduce the signal efficiency by less than 15$\%$. In the energy region below 0.5 GeV, the efficiency is greatly reduced, but even when using all available triggers, the efficiency is below 3$\%$. Therefore, we avoid the regions of lowest trigger efficiency by restricting our search to the range $\Amass$~$<$~8.97~$\Gmass$ and $\DMmass$~$<$~4.44~$\Gmass$.

We require exactly two oppositely charged tracks originating from the interaction point (IP) with impact parameters within $\pm$4.0~cm along the beam axis and 2.0~cm in the transverse plane. These two charged tracks are identified as pions by using the likelihood ratio $\mathcal{L_{\pi}}/(\mathcal{L}_{\pi}+\mathcal{L}_{K})$, where $\mathcal{L_{\pi}}$ and $\mathcal{L}_{K}$ are the likelihood with a pion and kaon hypothesis, respectively. The likelihood uses information from the central drift chamber (CDC), time-of-flight scintillation (TOF) counters, and aerogel Cherenkov counters (ACC), and the ratio is required to be larger than 0.6. To suppress contamination from electrons, we further require an electron identification, which is a similar likelihood ratio derived mainly from ECL information, to be less than 0.1. We estimate from MC that 90$\%$ of signal candidate events contain a pair of correctly reconstructed pions. Fake pions originate from muons ($<$ 6.4$\%$ of tracks), electrons ($<$ 3.8$\%$), and protons ($<$ 0.2$\%)$.  The tagged charged pion candidates are identified by using mainly CDC information due to the low transverse momenta of the tracks, hence resulting in higher fake rates in comparison to those in generic hadronic events triggered by Belle. The highest energy photon in the center-of-mass (CM) frame is chosen as the photon candidate in each event. This photon is required to have energy in the $\ups$ frame, $E_{\gamma}^{*}$, larger than 0.15~GeV, must lie in the polar angle range of $-$0.63~$<$~cos($\theta$)~$<$~0.84 of the ECL, must hit more than 2 calorimeter crystals, and have an energy-deposit ratio in 3$\times$3 over 5$\times$5 crystals around the shower center greater than 0.9.

The invariant recoil mass of the dipion system is defined as $M^{2}_{\text{recoil}}=s+M^{2}_{\pi\pi}-2\sqrt{s}\,E^{**}_{\pi\pi}$, where $\sqrt{s}$ = 10.02~$\Gmass$ is the $\upss$ resonance energy, $M_{\pi\pi}$ is the invariant mass of the dipion system, and $E^{**}_{\pi\pi}$ is the energy of the dipion system in the CM frame of the $\upss$.
The recoil mass is required to be between 9.450~$\Gmass$ and 9.475~$\Gmass$, which corresponds to the $\ups$ mass~\cite{Patrignani:2016xqp}. The vertex of the two pions is required to be near the IP with $\chi^{2}$/n.d.f.\ $<$ 11 from the vertex-constrained fit, and an opening angle of the dipion system in the $\ups$ frame larger than 45$^\circ$. The angle between the candidate photon and each charged track in the lab frame must satisfy cos($\theta_{\pi^{\pm}\gamma}$) $<$ 0.97 to reject photons due to bremsstrahlung and final-state radiation, while the azimuthal angle difference between the dipion system and the candidate photon must satisfy cos($\phi_{\pi\pi}-\phi_{\gamma}$) $>$ $-0.97$, to suppress QED background processes, such as $e^{+}e^{-}\rightarrow\gamma\pi^{+}\pi^{-}$. 

Neutral hadrons may pass these selections, thus we also require that the energy of the second-highest-energy photon in the CM frame and the remaining energy in the ECL both be less than 0.18~GeV.
To suppress events with a long-lived SM particle in the direction opposite to the candidate photon, we define the absolute azimuthal angle difference between a candidate photon and a candidate long-lived particle as $| 180^{\circ} - |\phi_{\gamma}-\phi_{\text{long}}||$, where the long-lived $K_L$ candidate is in the direction opposite to the photon. This absolute value of the angle difference is required to be larger than 20$^{\circ}$.
This selection rejects 54$\%$ of $\ups\rightarrow\gamma K_{L}K_{L}$, 98$\%$ of $\ups\rightarrow\gamma f_{2}'(1525)$, and 95$\%$ of $\ups\rightarrow\gamma f_{2}'(1270)$ events.

The selection criteria described above are optimized with the figure of merit S/$\sqrt{\text{B}}$, where S and B are the numbers of signal and expected background events, respectively, after applying all selections except the selection being evaluated. The signal efficiency ranges from 0.001$\%$ to 14$\%$ for the on-shell signal and from 0.0007$\%$ to 9.4$\%$ for the off-shell signal. The lowest efficiencies correspond to the highest $\Amass$ and $\DMmass$, respectively. The efficiency drop is due to the reduced trigger efficiency for low energy photons.

Irreducible background from the $\upss$ resonance is studied using a sample of 400 $\times$ 10$^{6}$ $\upss$ inclusive MC events, and categorized into three event types: tau-pair production $\upss\rightarrow\tau^{+}\tau^{-}$, leptonic decays $\ups\rightarrow l^{+}l^{-}$, and hadronic decays $\ups\rightarrow\gamma h h$. Taus can decay to charged pions and a tau neutrino, thus the slow charged pions in such decays can pass the selection criteria. Leptons $l$ and hadrons $h$ can escape the detector along the beam pipe, so that we only tag the two charged pions from the dipion transition and a photon. The leptonic decay backgrounds do not produce a peak in the $E_{\gamma}^{*}$ spectrum, but the hadronic decay backgrounds can produce such a peak. Both types of backgrounds peak in the $M_{\rm recoil}$ distribution. The background contributions from these backgrounds are predicted to be: 3.5~$\pm$~1.2 events from $\upss\rightarrow\tau^{+}\tau^{-}$ decays, 20.0~$\pm$~2.8 events from leptonic decays, and 1.2~$\pm$~0.7 events from hadronic decays. Continuum backgrounds are studied with an off-resonance data set collected about 60 MeV below the $\upsss$ resonance. This sample corresponds to an integrated luminosity of 40.41~fb$^{-1}$; we do not observe any significant peaking backgrounds. 

To search for a signal after the event selection, we use the two observables $M_{\text{recoil}}$ and $E_{\gamma}^{*}$. We construct probability density functions (PDFs) for signal and for background from the $\upss$ by using MC samples. Continuum background PDFs are created from $M_{\text{recoil}}$ sideband regions in the $\upss$ on-resonance data.
The recoil mass for the $\upss$ on-resonance is described with a double-sided Crystal Ball (CB) function~\cite{CBf}, and continuum in the recoil mass distribution is described with a second-order Chebyshev polynomial.
The bias in the $E_{\gamma}^{*}$ spectrum from the trigger efficiency is accounted for by multiplying the $E_{\gamma}^{*}$ PDFs by a parameterization of the trigger efficiency as a function of $E_{\gamma}^{*}$. 

The on-shell process $E_{\gamma}^{*}$ PDF is described with a CB function, and the off-shell process is described with a custom broad distribution function~\cite{Seong:2017iiq}. Each parameter of the $E_{\gamma}^{*}$ PDF is extracted
separately for the assumed values of $\Amass$ and $\DMmass$; these parameterized functions are used to search for a peak in the $E_{\gamma}^{*}$ spectrum. An exponential function is used for leptonic decay backgrounds, and a Gaussian function is used for hadronic decay backgrounds. Continuum backgrounds are described with the sum of an exponential function and a Gaussian function.
Tau-pair production from the $\upss$ does not peak either in the recoil mass distribution nor in the photon energy spectrum; therefore, we combine $\upss\rightarrow\tau^{+}\tau^{-}$ events and continuum backgrounds. The shape parameters of the recoil mass PDF are determined by using $\ups\rightarrow\mu^{+}\mu^{-}$ data.

We perform an unbinned extended log-likelihood fit in the two-dimensional ($M_{\text{recoil}}$, $E_{\gamma}^{*}$) space to estimate the yields of different event types. The fit is repeated for each possible signal mass value. We fix all shape parameters of the PDFs. Instead of floating three background yields, we combine the two $\ups$ background PDFs as $\mathcal{P}_{\ups} \propto
f_{ll} \,\mathcal{P}_{ll} + (1-f_{ll})\,\mathcal{P}_{hh}$, where $\mathcal{P}_{ll}$ and $\mathcal{P}_{hh}$ are the PDFs of the leptonic and hadronic decay backgrounds and $f_{ll}$ is the fraction of leptonic decay backgrounds, respectively. We use a fixed value of $f_{ll}=0.933\pm0.034$, obtained from the $\upss$ inclusive MC sample. To maximize the likelihood function and obtain signal yields, we vary two background yields and one signal yield, $N_{\text{cont}}$, $N_{\ups}$, and $N_{\text{sig}}$.

\begin{figure}[htb]
	\includegraphics[width=0.4\textwidth]{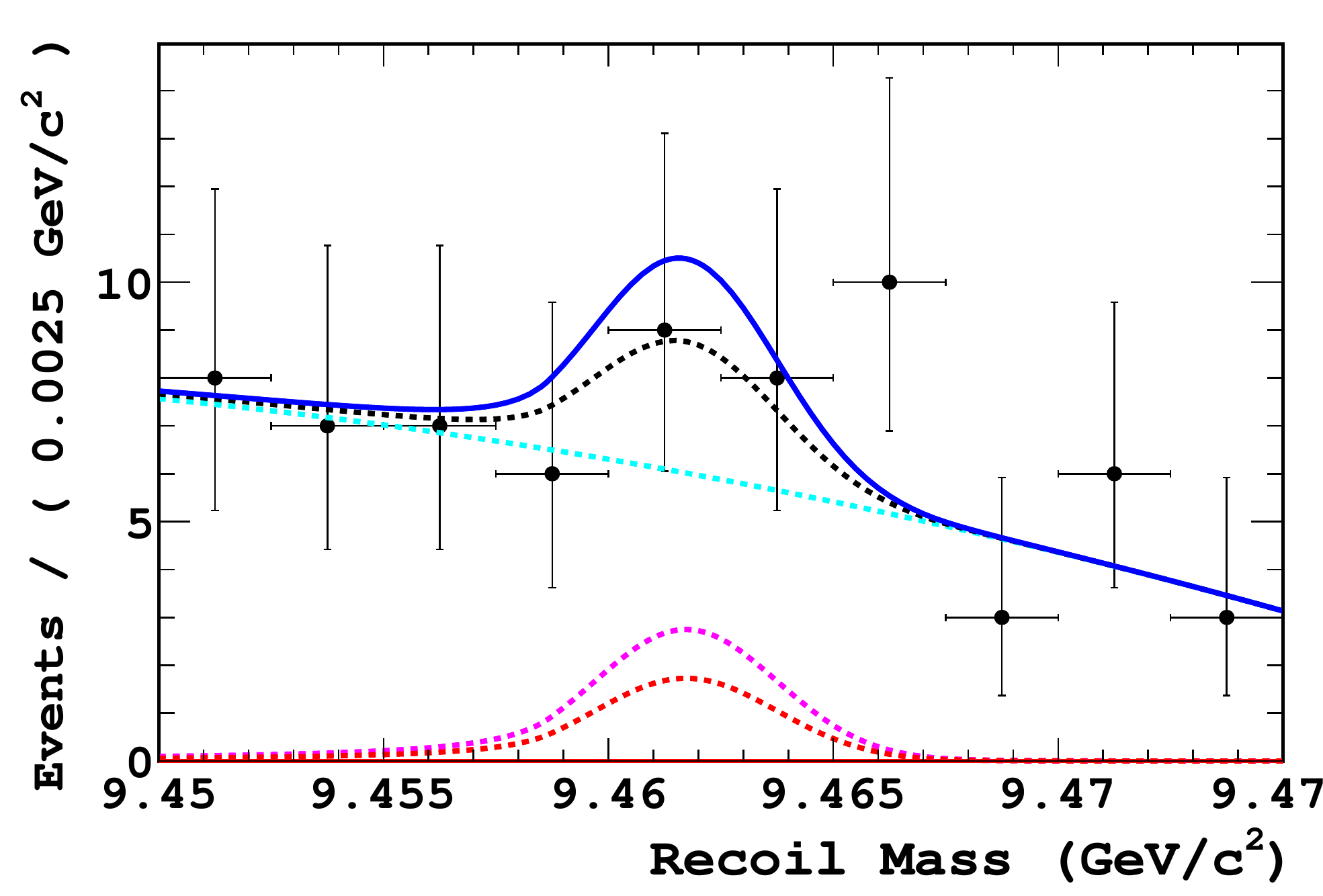}
	\includegraphics[width=0.4\textwidth]{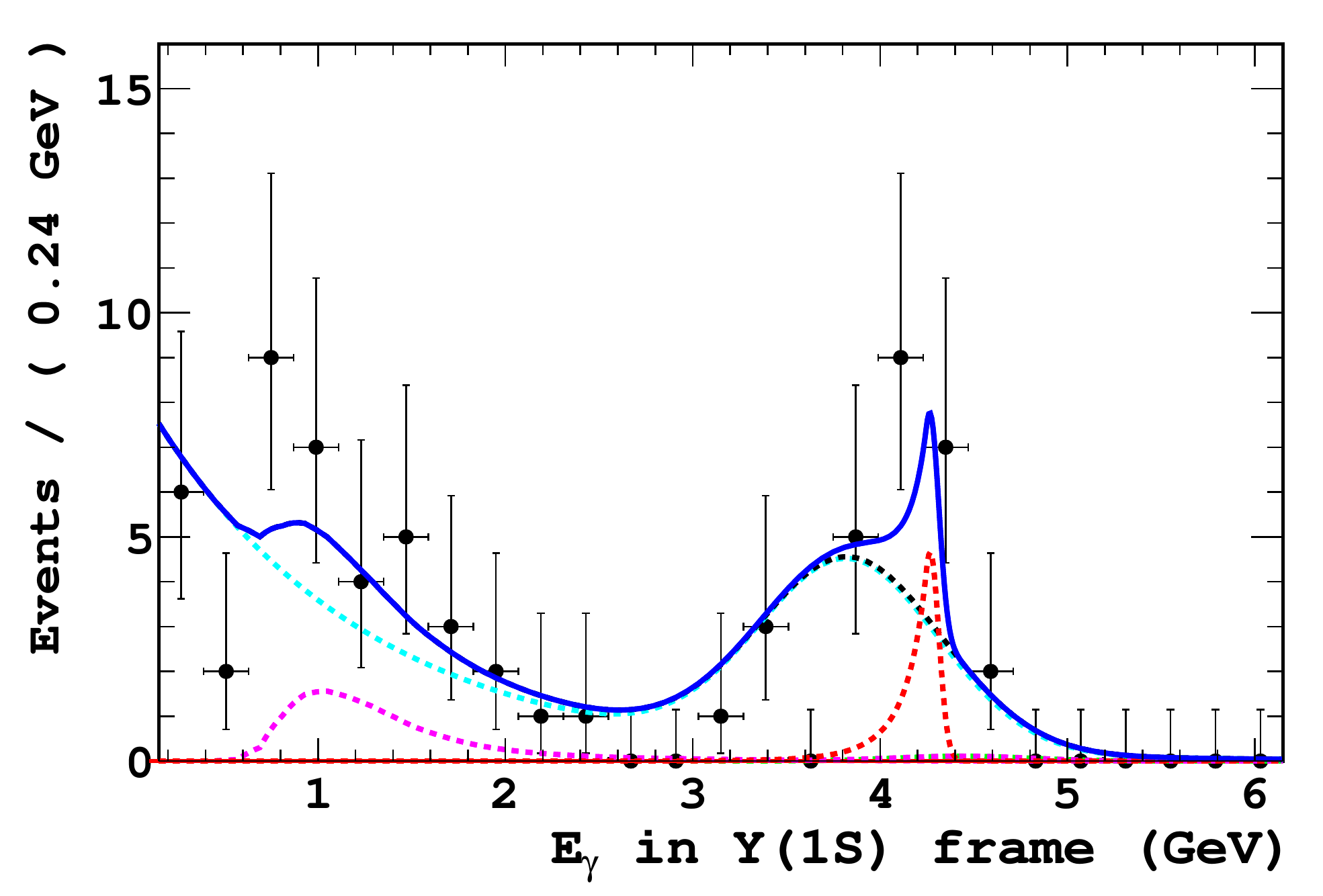}
	\caption{Fit result for the on-shell process mass scan point with $\Amass$ = 2.946~$\Gmass$, which has the highest local signal significance; 2.1$\sigma$. Top: $M_{\text{recoil}}$ distribution. Bottom: $E_{\gamma}^{*}$ distribution. The fitted components are continuum background (cyan dashed curve), $\ups$ decay background (magenta dashed curve), total background (black dashed curve), and the on-shell signal (red dashed curve). The blue solid curve shows the sum of all fitted components.  
	}
	\label{fig:fit}
\end{figure}

We search for a signal peak in the $E_{\gamma}^{*}$ and the $M_{\text{recoil}}$ distributions, in the mass ranges 0 $< M_{\A} <$ 8.97 GeV/c$^2$ (on-shell process) and 0 $< M_{\chi} <$ 4.44 GeV/$c^{2}$ (off-shell process) by repeating the extended log-likelihood fit for each value of $M_{\A}$ or $M_\chi$. For the on-shell case, we scan the photon energy in 353 steps that correspond to half the photon energy resolution, and step size in the range from 25 MeV to 4.0 MeV. For the off-shell case, we use 45 $M_{\chi}$ scan points with a fixed step size of 100 MeV. If the likelihood fit finds $N_{\text{sig}}$ $>$ 0, we compute the signal significance $S = \sqrt{2\text{ln}(\mathcal{L}_{\text{max}} / \mathcal{L}_{0})}$, where $-\ln\mathcal{L}_{\text{max}}$ is the negative log-likelihood value at the minimum and $-\ln\mathcal{L}_{0}$ is the minimum value for the background-only hypothesis.
We perform the mass scans and observe the largest local significance of 2.1$\sigma$ at $\Amass$ = 2.946~$\Gmass$; see Fig.~\ref{fig:fit}. And the largest local significance for the off-shell case is 1.4$\sigma$ at $\DMmass$ = 4.2~$\Gmass$. We observe no statistically significant signal and compute an upper limit (UL) at 90$\%$ C.L. on the signal yield ($N_{\text{UL}}$) by integrating the likelihood function $\int_{0}^{N_{\text{UL}}} \mathcal{L}(N_{\text{sig}})\, dN_{\text{sig}} = 0.9 \int_{0}^{\infty} \mathcal{L}(N_{\text{sig}})\, dN_{\text{sig}}$. The systematic uncertainty is accounted for in the limit calculation by convolving the likelihood with a Gaussian function, which has a width equal to the total systematic uncertainty. The upper limits (90$\%$ C.L.) on the BFs of the on-shell and the off-shell signals are then given by
$N_{\text{UL}}/(N_{\upss}\times\Br(\upss\rightarrow\ups\pi^{+}\pi^{-})\times\epsilon)$, where $\epsilon$ is the signal efficiency. 

Several sources of systematic uncertainties are included in the upper limits on the BFs. For most scan points, the observed yield $N_{\rm sig}$ is small, thus multiplicative signal uncertainties do not have a significant effect. The leading sources of systematic uncertainties are due to fit bias and PDF shape parameters.
The systematic uncertainty due to the BF of the dipion transition is estimated to be 1.46$\%$ based on Ref.~\cite{Patrignani:2016xqp}, and the uncertainty due to the number of $\upss$ events is 2.3$\%$. The uncertainty in the tracking efficiency for tracks with angles and momenta characteristic of signal events is about 1.4$\%$ per track.
The photon reconstruction contributes an additional 3.0$\%$ uncertainty.
Systematic uncertainties on the signal efficiency range from 0.2$\%$ to 0.7$\%$ at $\Amass$ $\leq$ 8.5~$\Gmass$ and from 0.7$\%$ to 30$\%$ at 8.5 $<$ $\Amass$ $<$ 8.97~$\Gmass$ for the on-shell signal; and from 0.3$\%$ to 0.8$\%$ at $\DMmass$ $\leq$ 4.0~$\Gmass$ and from 0.8$\%$ to 38$\%$ at 4.0 $<$ $\DMmass$ $<$ 4.44~$\Gmass$ for the off-shell signal. The uncertainty in the L1 trigger efficiency is estimated to be 13.5$\%$ by comparing the relative efficiency of the two L1 triggers in experiment against the same quantity in MC.
A possible bias in the fit is checked for by using toy MC samples, for the same values of $\Amass$ or $\DMmass$ used to generate the signal MC samples. For each signal mass, a toy MC is generated using the background and signal PDFs. The number of background events in the toy MC is obtained from the background-only fit to the $\upss$ on-resonance data and it is generated following the Poisson distribution. The signal yield is varied from zero to 11 events. For each signal mass and each signal yield, we generate 1000 toy MC events. We observe a fit bias of 0.001 for the on-shell signal and observe a bias that depends on $\DMmass$ for the off-shell signal. The largest fit bias for the off-shell process, 3.6 events, occurs at $\DMmass$ $\approx$ 3.5~$\Gmass$, which corresponds to a photon energy range 1~$< E_{\gamma}^{*} <$~2~GeV. In this range, the leptonic decay background influences the measured signal yield. Therefore, we assign a systematic uncertainty due to fit bias that varies with $\DMmass$ for the off-shell signal, and assign a 0.001 event systematic uncertainty for the on-shell signal.
Systematic uncertainties due to the PDF shapes are estimated by refitting with the shape parameters and the predicted $f_{ll}$ varied within their uncertainties. The continuum shape in the recoil mass distribution is also refit with a first-order Chebyshev polynomial function. We repeat the likelihood scan for each variation of this kind, and add in quadrature all of the resulting variations in the fitted yield at that signal mass. The largest systematic uncertainty of shape parameters is 2.5 and 2.8 events for the on-shell and the off-shell signal, respectively. We quote fit variation uncertainties depending on $\Amass$ and $\DMmass$.

\begin{figure}[htb]
     \includegraphics[width=0.4\textwidth]{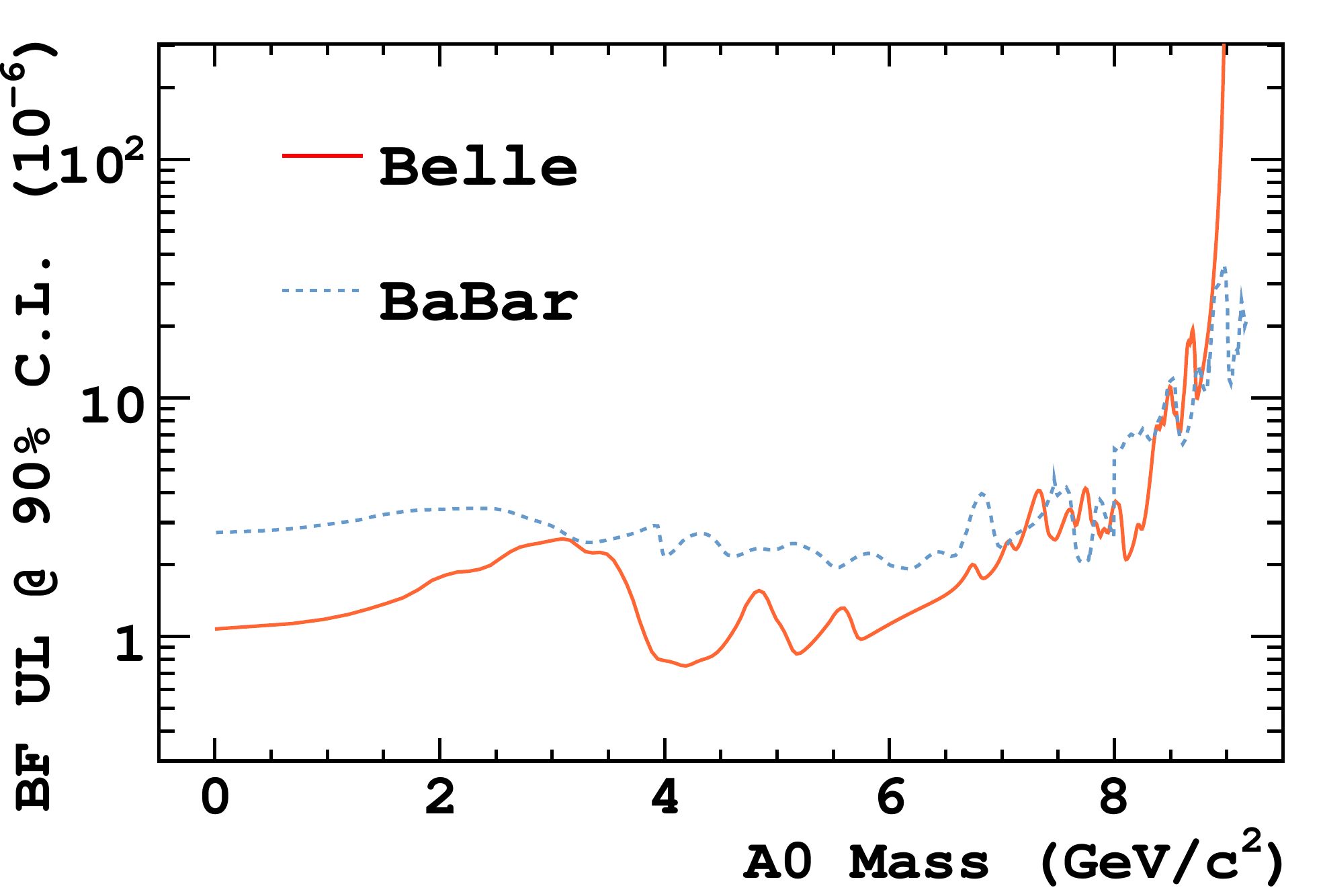}
    \includegraphics[width=0.4\textwidth]{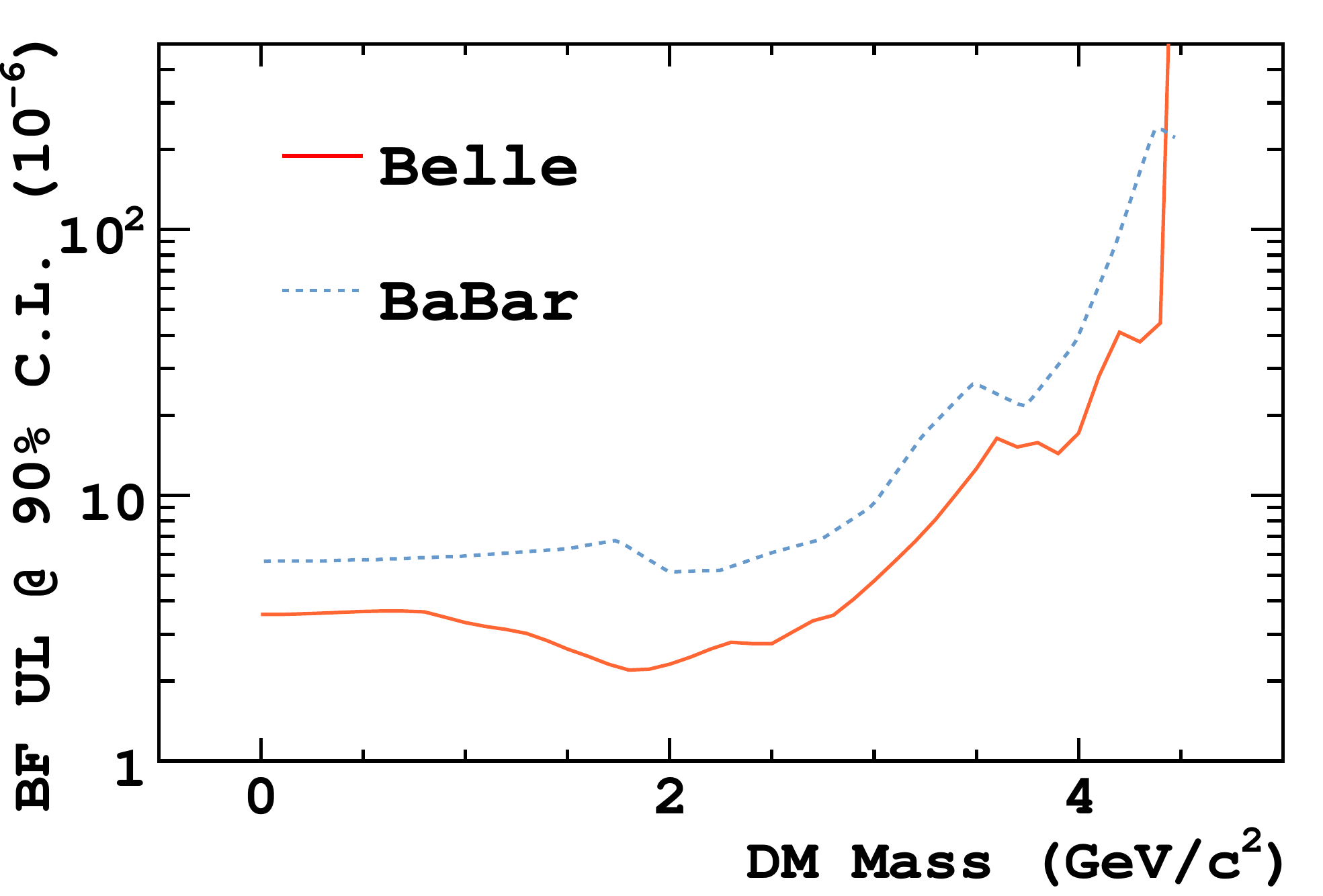}
	\caption{90$\%$ C.L. upper limits on the BFs of the on-shell process $\ups\rightarrow\gamma\A$ with $\A\rightarrow\chi\chi$ (top) and the off-shell process
$\ups\rightarrow\gamma\chi\chi$ (bottom). The orange solid curves are the Belle limits and the blue dashed curves are the BaBar limits. 
	}
	\label{fig:BF_limits}
\end{figure}

The estimated systematic uncertainty is included in the likelihood and we obtain the 90$\%$ C.L. upper limits on $\Br(\ups\rightarrow\gamma\A) \times\Br(\A\rightarrow\chi\chi)$ and $\Br(\ups\rightarrow\gamma\chi\chi)$ shown in Fig.~\ref{fig:BF_limits}. 
For the on-shell process, we achieve slightly better sensitivity in the low mass region than the BaBar result, and comparable or worse sensitivity in the high mass region. This low sensitivity is due to the lower trigger efficiency of Belle in that mass region. For the off-shell process, we achieve better limits than BaBar for all masses. Our limits are dominated by statistical uncertainties.

\begin{figure}[htb]
	\includegraphics[width=0.5\textwidth]{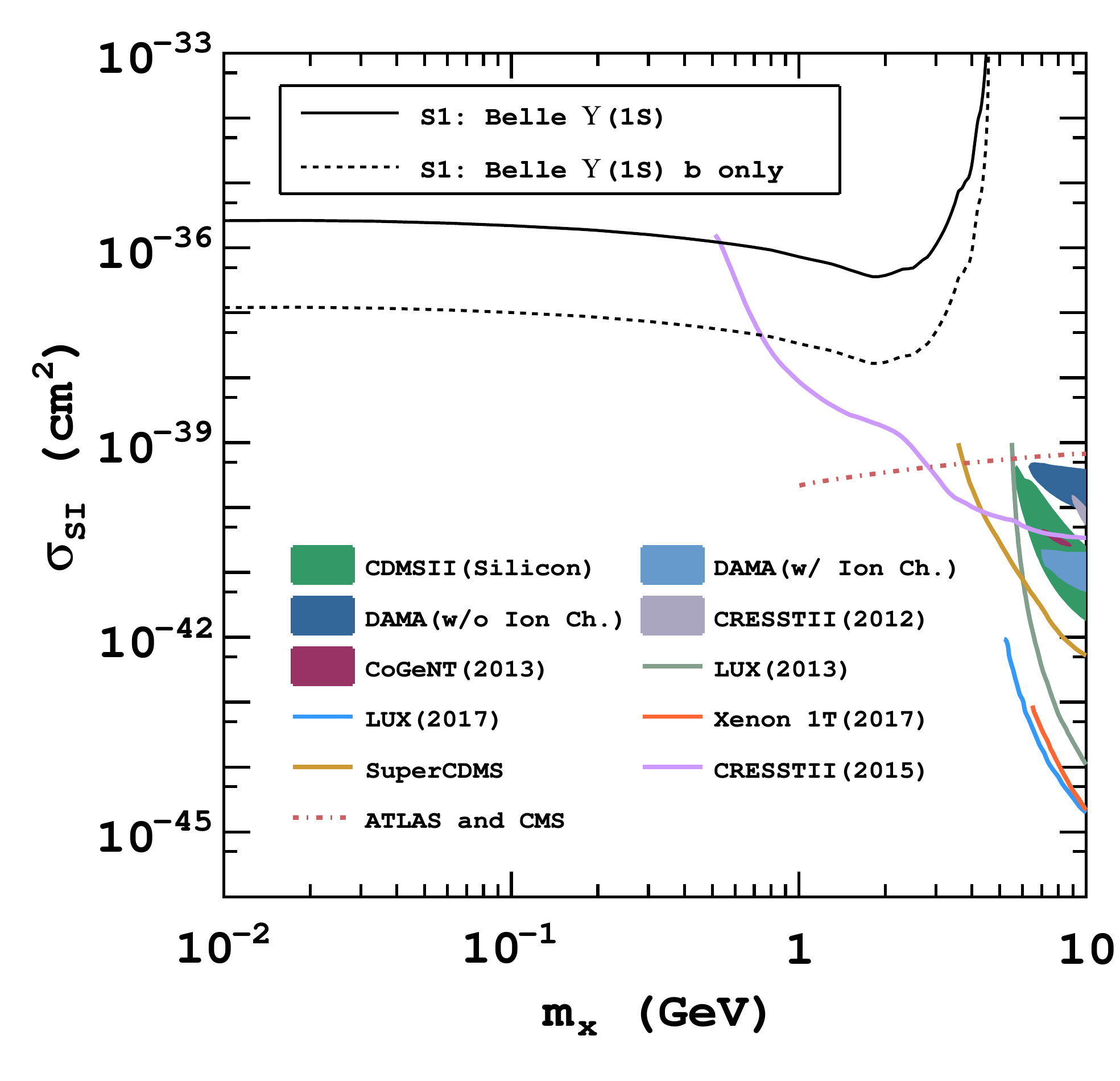}
	\caption{WIMP-nucleon spin-independent scattering cross-section limits at 90$\%$ C.L. The black solid and dashed curves are the upper limits obtained by assuming the WIMP couples to all quarks and only b-quarks, respectively. The 90$\%$ C.L. exclusion limits of LUX~\cite{Akerib:2016vxi}, CRESST II~\cite{Angloher:2015ewa}, SuperCDMS ~\cite{Agnese:2014aze}, and ATLAS~\cite{Aad:2014tda, Aad:2015zva} and CMS~\cite{Khachatryan:2014rwa, Khachatryan:2014rra} are shown for reference; and the 90$\%$ C.L. signal regions of CRESST II~\cite{Angloher:2011uu}, CoGeNT~\cite{Aalseth:2014jpa}, DAMA/LIBRA~\cite{Savage:2008er}, and CDMS II (Silicon)~\cite{Agnese:2013rvf} are also shown.
	}
	\label{fig:cross_section}
\end{figure}

The limit on the BF of the off-shell process can be converted into a WIMP-nucleon scattering cross section limit by using the procedure in Ref.~\cite{Fernandez:2015klv}. The off-shell process generated in this analysis corresponds to the S1 operator in Ref.~\cite{Fernandez:2015klv}, and we set new spin-independent (SI) WIMP-nucleon cross section limits, shown in Fig.~\ref{fig:cross_section}. We place one set of limits assuming that the WIMP couples to all quarks, and another set of limits assuming it couples to b-quarks only. These limits extend down into the interesting low-mass WIMP region unreachable by currently running direct detection
experiments. It should be noted that these limits are valid regardless of whether the CP-odd light Higgs exists, but they do assume the existence of some new spin-zero boson, because the S1 operator is used to set the limit.


To conclude, we have performed the first Belle search for the on-shell process, $\ups\rightarrow\gamma\A$ with $\A\rightarrow\chi\chi$, and the off-shell process, $\ups\rightarrow\gamma\chi\chi$, and have set upper limits on the branching fractions at 90$\%$ C.L. in the mass ranges 0~$< M_{\A} <$~8.97~$\Gmass$ and 0 $< M_{\chi} <$ 4.44~$\Gmass$. Our results improve on the existing limits from BaBar, mainly for the off-shell case. We have used the Belle branching fraction limit on the off-shell process to set new limits on the SI WIMP-nucleon scattering cross section. We uniquely constrain the low mass dark matter region where direct detection experiments do not yet have sensitivity, under the general assumption that a new spin-zero boson exists. We expect that this work can be extended significantly in the near future by using data from the Belle II experiment, which is currently being commissioned~\cite{Lewis:2018ayu}, and by searching for WIMPs assuming other contact operators, as discussed in Ref.~\cite{Fernandez:2015klv}.

We thank the KEKB group for excellent operation of the accelerator; the KEK cryogenics group for efficient solenoid operations; and the KEK computer group, the NII, and PNNL/EMSL for valuable computing and SINET5 network support. We acknowledge support from MEXT, JSPS and Nagoya's TLPRC (Japan); ARC (Australia); FWF (Austria); NSFC and CCEPP (China); MSMT (Czechia); CZF, DFG, EXC153, and VS (Germany); DST (India); INFN (Italy); MOE, MSIP, NRF, RSRI, FLRFAS project and GSDC of KISTI (Korea); MNiSW and NCN (Poland); MSHE under contract 14.W03.31.0026 (Russia); ARRS (Slovenia); IKERBASQUE and MINECO (Spain); SNSF (Switzerland); MOE and MOST (Taiwan); and DOE and NSF (USA).


%

\end{document}